\documentclass[fleqn,twoside]{article}
\usepackage{espcrc2}
\usepackage[dvips]{graphicx}
\usepackage{pstricks}

\title{Fractal Structure of 4D Euclidean Simplicial Manifold}

\author{H. S. Egawa
        \address{Research Institute of Educational Development, 
                 Tokai University,
                 2-28-4 Tomigaya, Shibuya, Tokyo 151-8677 Japan}
        ,
        S. Horata
        \address[SOKENDAI]{Coordination Center for Research and Education,
                           The Graduate University for Advanced Studies,
                           Hayama, Kanagawa 240-0193 Japan}
        \address{Institute of Particle and Nuclear Studies, KEK,
                 High Energy Accelerator Research Organization,
                 1-1 Oho, Tsukuba, Ibaraki 305-0801 Japan}
        \thanks{presented by S. Horata}
        and
        T. Yukawa
        \addressmark[SOKENDAI]
}
\begin{document}

\begin{abstract}
The fractal properties of four-dimensional Euclidean simplicial 
manifold  generated by the dynamical triangulation
are analyzed on the geodesic distance $D$ between two vertices instead of
the usual scale between two simplices. 
In order to make more unambiguous measurement of the fractal dimension,
we employ a different approach from usual, by measuring the box-counting
dimension which is computed by counting the number of spheres with the 
radius D within the manifold. 
The numerical result is consistent to the result of the random walk model 
in the branched polymer region. We also measure the box-counting dimension 
of the manifold with additional matter fields. Numerical results suggest 
that the fractal dimension takes value of slightly more than 4 near the 
critical point.
Furthermore, we analyze the correlation functions as functions of the 
geodesic distance. 
Numerically, it is suggested that the fractal structure of four-dimensional 
simplicial manifold can be properly analyzed in terms of the distance 
between two vertices. 
Moreover, we show that the behavior of the correlation length regards the 
phase structure of 4D simplicial manifold.
\end{abstract}

\maketitle

\section{Introduction and model}
Recently, Euclidean simplicial quantum gravity(SQG) with the additional 
matter fields has been investigated with using the Monte-Carlo simulation.
From the previous numerical simulation\cite{PHASE,MATDEP}, we expect
the four-dimensional(4D) simplicial model realizes the realistic 4D quantum 
gravity.
Especially, in Ref.~\cite{MATDEP}, it is shown that the relation between
SQG and the 4D conformal gravity\cite{BMI}, calculating with the
grand-canonical method.

However, in order to discuss the relation between the discretized model 
and the continuum theory in detail, we need to analyze the correlations 
on the 4D simplicial manifold ${\cal M}_{\rm DT}({\rm 4D})$.
Namely, we need to discuss the geometry on ${\cal M}_{\rm DT}({\rm 4D})$
instead of the statistical properties which have been discussed until now.
%
%
%

In order to discuss the geometry on ${\cal M}_{\rm DT}({\rm 4D})$,
we analyze the configurations of the 4D SQG coupled with $N_A$ U(1) gauge 
fields and $N_X$ massless scalar fields.
%
%
The partition function $Z$ is given as the sum of the all of the
configuration of 4-simplices $T$, 
\begin{eqnarray}
Z = \sum_{T} e^{-S_{EH}(T)} \prod_{N_A,N_X} \qquad \qquad \qquad \qquad \\ \nonumber 
\left(\int \prod_{l \in T} dA_l e^{-S_{A}(A_l)} 
\int \prod_{i \in T} dX_i e^{-S_{X}(X_i)} \right),
\end{eqnarray}
where $S_{EH}$, $S_{X}$ and $S_{A}$ is the action for the gravity, 
the scaler field $X_i$ on the vertex $i$ and the vector field $A_{l}$ on
the link $l$ respectively.
For the action of the gravity part, we use the discretized Einstein-Hilbert 
action,
\begin{equation}
S_{EH}[\kappa_{2},\kappa_{4}]=\kappa_{4}N_{4}-\kappa_{2}N_{2}
\end{equation}
where $N_i$ denotes the number of the $i$-simplices, $\kappa_2$ is related to
Newton's constant and $\kappa_{4}$ is related to Cosmological constant.
The action for the scalar field $X_i$ on the vertex $i$ becomes
\begin{equation}
S_X = \sum_{ij} o(l_{ij}) (X_i - X_j)^2,
\end{equation}
where $o(l_{ij})$ is number of 4-simplices sharing link $l_{ij}$,
Then, the action of vector field $A_{ij}$ on link $l_{ij}$
\begin{equation}
S_A = \sum_{ijk} o(t_{ijk}) (A(l_{ij})+A(l_{jk})+A(l_{ki}))^2,
\end{equation}
where $o(t_{ijk})$ is number of 4-simplices sharing triangle $t_{ijk}$.

In order to attempt the canonical simulation, we add the volume fixing term
$\delta S$ to the total action,
\begin{equation}
\delta S(N_4, V_4) = \delta \kappa_4 \left( N_4 - V_4 \right)^2, 
\end{equation}
where $\delta \kappa_4$ is the parameter to fix the volume $N_4$ and 
$V_4$ denotes the size of the target volume.
At $N_4 = V_4$, we take a sample of ${\cal M}_{\rm DT}({\rm 4D})$ and
attempt to measure.

\section{Fractal dimension based on the geodesic distance}

In order to analyze the geometry of ${\cal M}_{\rm DT}({\rm 4D})$,
we discuss the fractal properties based on the geodesic distance with
measuring the fractal dimension and the correlation functions of the
 additional fields.

We consider some different definitions of the geodesic distance on
${\cal M}_{\rm DT}({\rm 4D})$.
One of such the definitions is the shortest distance between
two 4-simplices along the 4-simplices, so-called the 
simplex-simplex geodesic distance. 
%
%
However, we expect that the expansion of the volume along the
simplex-simplex geodesic distance is restricted, 
since the number of 4-simplices boarding on the 4-simplex is fixed.
Another definition is the shortest distance between two vertices along
the link, so-called the vertex-vertex geodesic distance.
%
%
%

In order to discuss the difference between the two distances, we
introduce the box-counting dimension 
${\rm dim}_{\rm BC}({\cal M}_{\rm DT})$ on ${\cal M}_{\rm DT}$,
\begin{equation}
{\rm dim}_{{\rm BC}}({\cal M}_{\rm DT}) = -\frac{\log (N(S(r(D)))}{\log r(D)},
\end{equation}
where $D$ denotes the scale of the geodesic distance, $N(S(r(D)))$ denotes the
number of the closed-surface $S(r(D))$ with the radius $r(D)$.
The box-counting method is known as one of the standard methods to
define the fractal dimension.

\begin{figure}
\centerline{\resizebox{70mm}{55mm}{\includegraphics{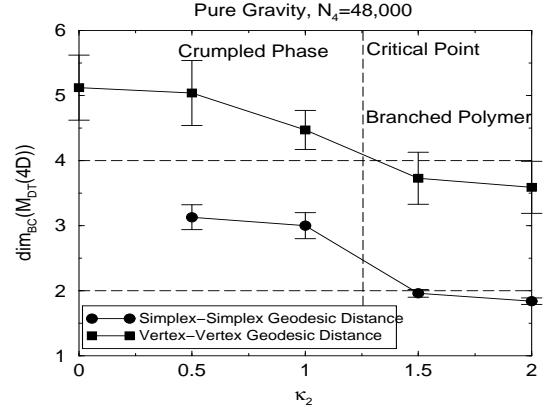}}}
\vspace{-8mm}
\caption{
The box-counting dimensions based on each geodesic distances versus
the coupling constant $\kappa_2$ in the case of pure gravity with
$V_4=48,000$. 
}
\label{fig.BC}
\end{figure}
In Figure~\ref{fig.BC}, we show the 
${\rm dim}_{\rm BC}({\cal M}_{\rm DT}({\rm 4D}))$ based on each geodesic 
distances in the case of pure gravity.
The numerical results suggest that each geodesic distance need rescaling
in order to be equivalent to the other definition of geodesic distance.
We consider that this rescaling factor is caused by the difference of
the degree of freedom of each distances on the discretized manifold.

From the behavior of the box-counting dimension in the branched polymer
region, the simplex-simplex geodesic distance is the intrinsic scale and
that it needs the rescaling factor for agreement with the unique shortest 
distance on the manifold.
On the contrary, numerical results suggest that the vertex-vertex geodesic
distance doesn't need the extra rescaling factor and that it is close to 
the unique shortest distance.
Then, we choose the vertex-vertex geodesic distance for analysis of the
geometry of ${\cal M}_{\rm DT}(4D)$.

\section{Behavior of the correlation function}
\begin{figure}
\centerline{\resizebox{70mm}{55mm}{\includegraphics{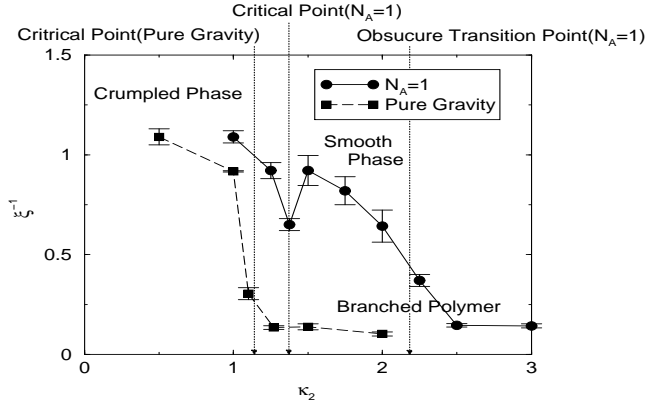}}}
\vspace{-10mm}
\caption{
The correlation length for each coupling constant $\kappa_2$ with 
$V_4=32,000$. 
}
\label{fig.col}
\end{figure}

We analyze the correlation functions on the vertex-vertex geodesic
distance instead of the conventional geodesic distance of
simplex-simplex distance.
In order to analyze the behavior of the correlation function, we introduce
the correlation length $\xi$ for the operator ${\cal O}(x)$ on the vertex 
$x$,
\begin{equation}
\xi^{-1} = - \frac{\log (< {\cal O}(x) {\cal O}(y) > )}{|x-y|}.
\end{equation}
This value diverges to infinity at the critical point of second order phase
 transition.

In Figure~\ref{fig.col}, we show the numerical results in the case of the pure
gravity and the additional one vector field $N_A=1$.
The numerical result suggests that the value of $\xi$ indicates the property 
of manifold and the phase structure.
Moreover, from the finite-size-scaling of the value of $\xi$  at the critical point between the crumpled phase and smooth phase in the case of $N_A=1$,
$\xi$ diverges to infinity for the large $N_4$ limit.

While the divergence of $\xi$ is not found both at the critical point of
the pure gravity case and at the obscure transition point between the
smooth phase and the branched polymer.
The numerical results suggest that the type of transition at the
critical point of the additional matter case is different from the type
of both the critical point of the pure gravity and the obscure transition
point.
%
%

\begin{figure}
\centerline{\resizebox{70mm}{55mm}{\includegraphics{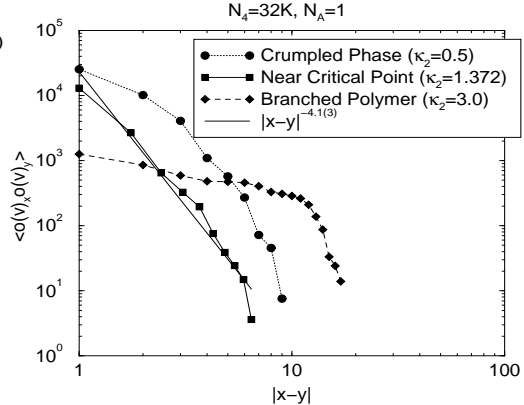}}}
\vspace{-10mm}
\caption{
The correlation functions for $o(v)$, which is the number of
vertices sharing a simplex.
}
\label{fig.TF_NA=1}
\end{figure}

Actually, we found the power low behavior of the correlation functions
at the critical point in the case of $N_A=1$. 
In Figure~\ref{fig.TF_NA=1}, we show the correlation functions for the local
scalar curvature.
The deviation from the power low can be considered as the finite size effect.
%
%
%

\section{Summary and Discussion}
In order to discuss the geometry on ${\cal M}_{\rm DT}({\rm 4D})$, 
we analyze the geodesic distances defined along the lattice building 
blocks.
From the numerical result of the box-counting dimension, 
the simplex-simplex geodesic distance need extra in order to be equivalent
to the vertex-vertex geodesic distance.
Moreover, we show the behavior of the correlation function numerically.
The correlation length realizes the phase diagram and its peak value 
diverges to the infinity at the critical point for the large $N_4$ limit.
Then, the numerical results of the box-counting dimension and the
correlation function suggest that the massless mode propagates on
the simplicial manifold which has the fractal dimension nearly 4.

In conclusion, the geometry on the simplicial manifold can be analyzed
on the basis of the vertex-vertex geodesic distance.

\end{document}